\documentstyle[sprocl,psfig]{article}

\def\be{\begin{equation}}
\def\ee{\end{equation}}
\def\bea{\begin{eqnarray}}
\def\eea{\end{eqnarray}}

\def\bm#1{%
\mathchoice
{{\hbox{\boldmath$\displaystyle#1$\unboldmath}}}%
{{\hbox{\boldmath$\textstyle#1$\unboldmath}}}%
{{\hbox{\boldmath$\scriptstyle#1$\unboldmath}}}%
{{\hbox{\boldmath$\scriptscriptstyle#1$\unboldmath}}}%
}

\begin{document}

\title{TWO PION PRODUCTION IN NUCLEON-NUCLEON COLLISIONS AT INTERMEDIATE
ENERGIES}

\author{L. ALVAREZ-RUSO$^a$, E. OSET$^a$, E. HERN\'ANDEZ$^b$}

\address{$^a$Departamento de F\'{\i}sica Te\'orica and IFIC,
Centro Mixto Universidad de Valencia-CSIC, 46100 Burjassot (Valencia), Spain\\
$^b$Grupo de F\'{\i}sica Nuclear, Facultad de Ciencias, 
Universidad de Salamanca,\\37008 Salamanca, Spain} 

\maketitle\abstracts{We have developed a model for the 
$N N \rightarrow N N \pi \pi$ reaction and evaluated cross sections for the 
different charged channels. The low energy part of those channels where the 
pions can be in an isospin zero state is dominated by $N^* (1440)$ excitation, 
driven by an isoscalar source followed by the decay 
$N^* \rightarrow N (\pi \pi)^{T=0}_{S-wave}$. 
At higher energies, and in channels where the pions are not in T=0, $\Delta$ 
excitation mechanisms become relevant.}

Pion production in $NN$ collisions is one of the sources of information on
nucleon-nucleon interaction and resonance properties. 
Nowadays, two pion production in $p p$ collisions is a subject of 
experimental research at the CELSIUS storage ring by the 
WASA/PROMICE collaboration~\cite{jan}. A direct comparison of our theoretical 
results with the measured total cross sections and invariant-mass 
distributions will certainly provide useful information about the mechanisms 
governing this reaction. 

Our model can also be relevant to understand certain features observed
in other two pion production reactions; namely, 
$n\, +\, p \rightarrow d\, +\, (\pi \pi)^0$ and
$p\, +\, d \rightarrow {^3}He +\, \pi^+ \, \pi^-$.
The first of them was experimentally studied at LAMPF for a neutron beam
energy of 800 MeV~\cite{holas}, and is also an important ingredient of the
deuteron spectra measured with a 1160 MeV neutron beam at 
Saturne~\cite{plouin}, where the ABC effect was first observed in 
the free nucleon reaction. The second is being studied at COSY in the MOMO 
experiment~\cite{momo}.   

Finally, the present model has also repercussions for the reaction
$p\, +\, p \rightarrow p\, +\, p \, + \,\pi^0$ since a possibly relevant 
mechanism can be obtained from
$p \, +\, p \rightarrow p\, +\, p \, + \,\pi^0 \pi^0$ when one of the
$\pi^0$'s is emitted and the other is absorbed~\cite{eli}. 

In order to build the model, we have closely followed the guidelines of a 
previous model for the $\pi N \rightarrow \pi \pi N$ reaction~\cite{manolo}. 
A relevant finding of that work is that, even at threshold, the Roper 
resonance excitation and subsequent decay into $N(\pi \pi)^{T=0}_{S-wave}$ is 
a very important mechanism. In our case, the situation is similar, but more 
involved because one has to deal with the  $NN \rightarrow N N ^*$ transition, 
which is poorly known. Below, we present a brief description of the ingredients 
of the model. Details can be found in Ref. 7.

\begin{figure}[h]
\centerline{\psfig{file=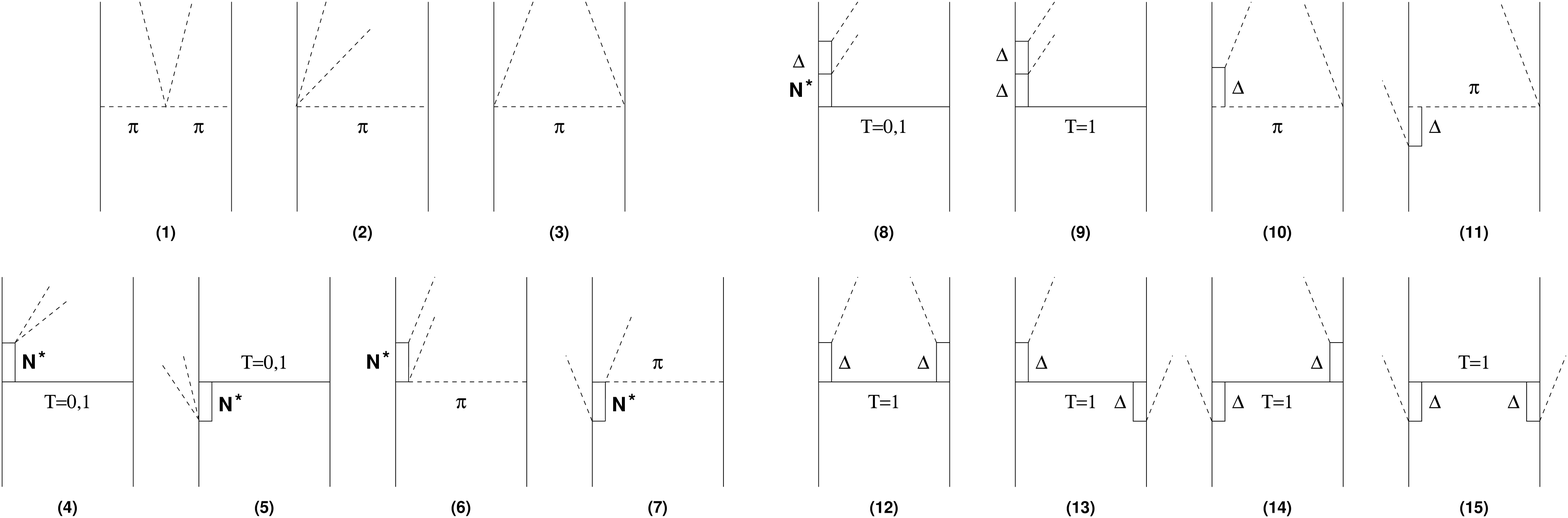,height=4.5cm,width=11.cm}}
\caption{Complete set of Feynman diagrams of our model.}
\label{fig1}
\end{figure}

Diagrams (1), (2) and (3) of Fig. 1 can be derived from the
lowest order SU(2) chiral Lagrangians containing both pions and nucleons
at tree level

\begin{equation}
{\cal L} = {\cal L}_2 + {\cal L}_1^{(B)} \,.
\end{equation}
The expressions for these Lagrangians are given, for instance, in Ref 8. 
With these ingredients we generate only the isovector part of the s-wave 
$\pi N$ amplitude, needed for diagram (3). The much smaller isoscalar part 
appears at higher order. We just consider it empirically

\begin{equation}
{\cal L} = -4 \pi \frac{\lambda_1}{m_\pi} \bar{\psi} \bm{\phi}^2 \psi \, ,
\quad \lambda_1 = 0.0075 \,.
\end{equation}
In spite of the fact that these amplitudes do not vanish at threshold, their
contribution to the total cross section is very small, even at
threshold, in most channels ( see Fig. 2 ).

In order to obtain the amplitudes for diagrams (9)-(15) we need the
following phenomenological Lagrangians

\begin{equation}
{\cal L}_{\Delta N \pi} = \frac{f^*}{m_\pi} \psi^{\dagger}_\Delta 
S^{\dagger}_i (\partial_i\phi^\lambda) T^{\dagger \lambda} \psi_N \, + \, h. c.
\end{equation}
\begin{equation}
{\cal L}_{\Delta \Delta \pi} = \frac{f_\Delta}{m_\pi} \psi^{\dagger}_\Delta
S_{\Delta i} (\partial_i\phi^\lambda) T_{\Delta}^{\lambda} \psi_N \, + \, h. c.
\end{equation}
where $S^\dagger$ ($T^\dagger$) and $S_\Delta$ ($T_\Delta$) are the spin
(isospin) 1/2 $\rightarrow$ 3/2 transition operator and spin (isospin) 3/2
operator respectively. The coupling constant
$f^*$= 2.13 is obtained from the $\Delta \rightarrow N \pi$ partial width
~\cite{pdg} and  $f_\Delta$ = 4/5 $f_{N N \pi}$ comes from SU(6) quark model.
Our T = 1 exchange potential includes $\pi$ and $\rho$ exchange and short 
range correlations~\cite{weise}. All these terms are quite small except 
diagram (12), which becomes relevant at energies $T_p >$ 1 GeV.   

We use the information about the $N^*(1440)$ properties contained in the PDG 
book~\cite{pdg} in order to extract the couplings required for diagrams (4)-(8):

\begin{equation}
{\cal L}_{N^* N \pi}
= \frac{\tilde{f}}{m_\pi} \psi^{\dagger}_{N^*} \sigma_i (\partial_i
\bm{\phi}) \bm{\tau} \psi_N  \, + \, h. c.
\end{equation}
\begin{equation}
{\cal L}_{N^* \Delta \pi} = \frac{g_{N^* \Delta \pi}}{m_\pi}
\psi^{\dagger}_{\Delta} S^{\dagger}_i (\partial_i \phi^{\lambda}) T^{\dagger
\lambda} \psi_{N^*} \, + \, h. c.
\end{equation}
Here, the couplings $\tilde{f} = 0.477$ and $g_{N^* \Delta \pi} = 2.07$ are
obtained from the corresponding partial widths assuming branching ratios of
65 \% and 25 \% respectively, and a total width of 350 MeV. 

The Lagrangian for the $N^* \rightarrow N (\pi \pi)^{T=0}_{S-wave}$ decay 
channel is given in Ref. 12. After expanding in pion fields one
gets 

\begin{equation}
{\cal L}_{N^* N \pi \pi} = - c^*_1 \frac{m^2_\pi}{f^2} \bar{\psi}_{N^*}
\bm{\phi}^2 \phi_N - c^*_2 \frac{1}{f^2 {M^*}^2} (\partial_\mu \partial_\nu 
\bar{\psi}_{N^*})(\bm{\tau} \partial_\mu \bm{\phi})
(\bm{\tau} \partial_\nu \bm{\phi})\psi_N \, + \, h. c.
\end{equation}
In this case, the free parameters $c^*_1$ and $c^*_2$ can not be both
obtained from the partial decay width. They can just be constrained to an
ellipse~\cite{meissner}. In order to further constrain these parameters we use 
the model of Ref. 6 for the reaction 
$\pi^- p \rightarrow \pi^+ \pi^- n$. The best agreement with the experiment is 
obtained with $c^*_1 = -7.27\, GeV^{-1}$ and $c^*_2 = 0$ ( Set I ) but the 
experimental errors are also compatible with $c^*_1 = -12.7\, GeV^{-1}$ and 
$c^*_2 = 1.98\,GeV^{-1}$ ( Set II).

In this case, the T=1 exchange is similar to the one for the $\Delta$ terms,
but in addition we must consider an exchange in the T=0 channel. In a
recent analysis of the ($\alpha, \alpha')$ reaction on a proton target
~\cite{hirenzaki}, the strength of the isoscalar $N N \rightarrow N N^*$
transition was extracted by parameterizing the transition amplitude in terms of
an effective ``$\sigma$'', which couples to $N N$ as the Bonn model $\sigma$
~\cite{bonn}

\begin{equation}
{\cal L}_{\sigma N N}  = g_{\sigma N N} \bar\psi \phi \psi, \quad
g^2_{\sigma N N}/ 4 \pi = 5.69 
\end{equation}
and couples to $N N^*$ 
\begin{equation}
{\cal L}_{\sigma N N^*}  = g_{\sigma N N^*} \bar\psi_{N^*} \phi \psi \, + \,
h.c.
\end{equation}
with a strength provided by a best fit to the data: 
$g^2_{\sigma N N^*}/ 4 \pi$ = 1.33, and such that $g_{\sigma N N}$ and 
$g_{\sigma N N^*}$ have the same sign.

The amplitudes corresponding to diagrams (4), (5) give a non-vanishing 
contribution at threshold, which is by far the dominant one below 1 GeV for 
those channels where the pions can be in T=0. The contribution of the T=0 
exchange is an order of magnitude larger than the T=1 exchange, which is 
diminished due to the short range correlations.

\begin{figure}[h]
\begin{minipage}{.47\linewidth}
\centerline{\psfig{file=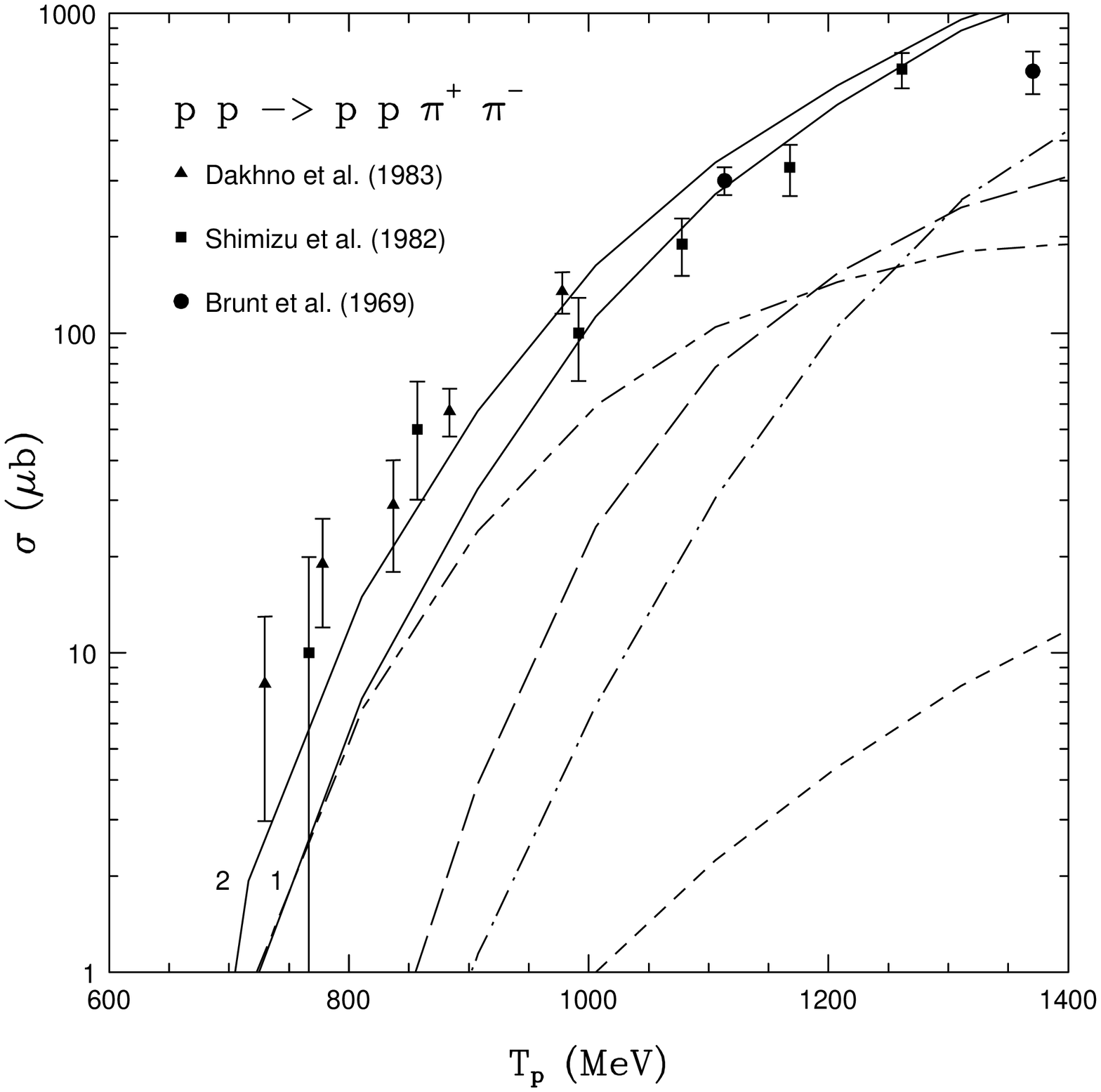,height=5.cm,width=5.cm}}
\end{minipage}
\hfill
\begin{minipage}{.47\linewidth}
\centerline{\psfig{file=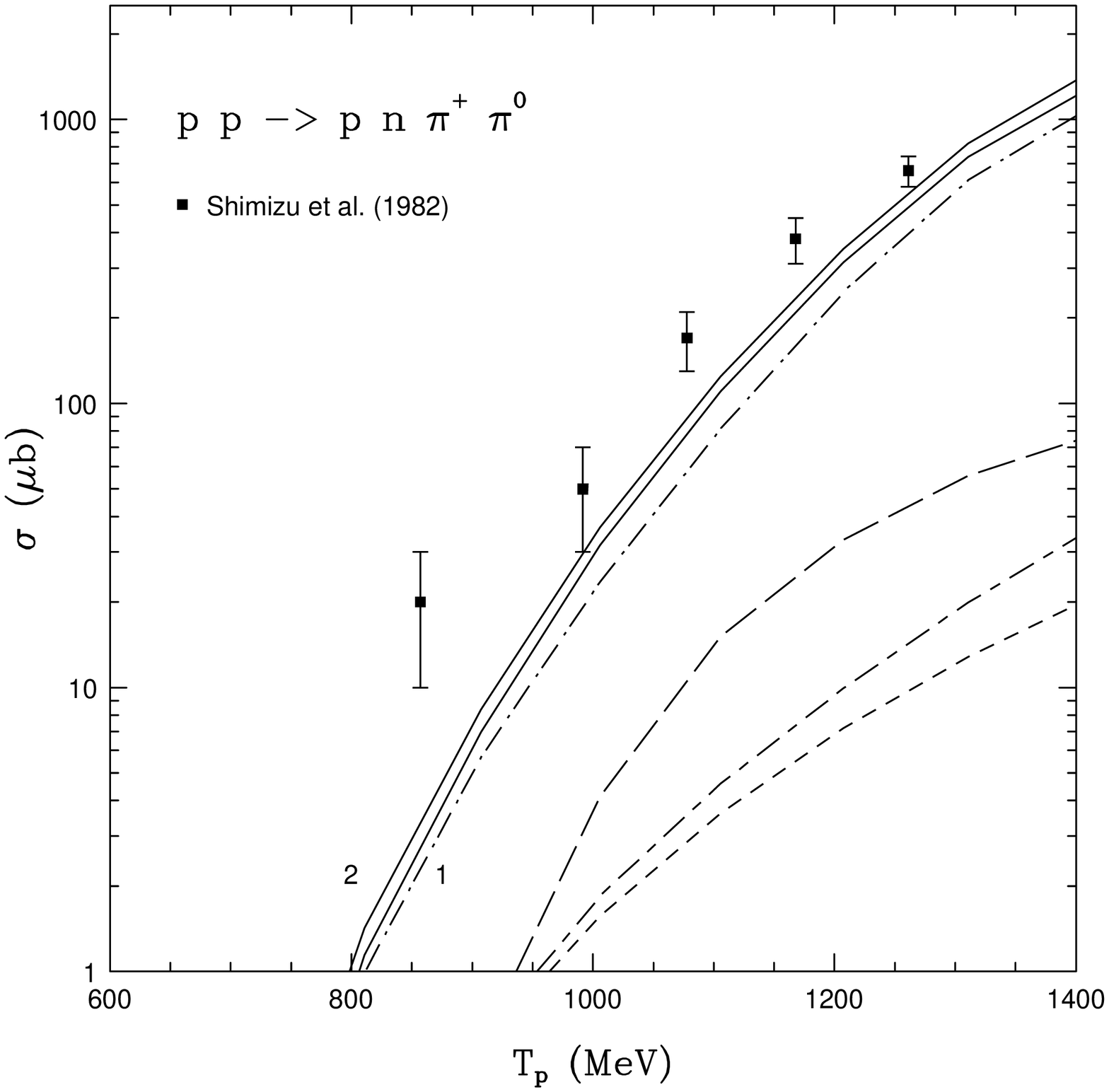,height=5.cm,width=5.cm}}
\end{minipage}
\caption{Total cross sections for two of the channels, as a function of the
incoming proton kinetic energy in lab. frame. Solid line, total ( line
labelled 1 for set I and 2 for set II ); long-short dashed line, 
$N^* \rightarrow  N (\pi \pi)^{T=0}_{S-wave}$; long-dashed line, 
$N^* \rightarrow \Delta \pi$; dash-dotted line, $\Delta$ excitation 
mechanisms; short-dashed line, non-resonant terms from diagrams (1)-(3). 
The partial contributions are calculated with set I. Experimental data are 
taken from Ref. 14.}
\label{fig2}
\end{figure}

We implement the final state interaction (FSI) in a simplified version of the
model, meant to work at $T_p < 900$ MeV and for the
$p p \rightarrow p p \pi^+ \pi^-$, where we can assume that the total cross
section is clearly dominated by the $N^*$ excitation driven by an isoscalar
source. Since the energy of the incoming nucleons must be
high enough to produce two pions, we neglect initial state interaction.      
For the configuration of the figure, the amplitude is

\begin{equation}
{\cal M}^{fsi} = \int \frac{d \bm{q}}{(2 \pi)^3} {\cal M}^{free} (q,p_{N^*})
\tilde{\varphi}_{\bm{k}} ( \bm{P} )
\end{equation}
where 
\begin{equation}
\bm{k} = \frac{\bm{p_4} - \bm{p_3}}{2}\, ,  \quad \bm{P} = \bm{q} +
\frac{\bm{p_5} + \bm{p_6} + \bm{p_2} - \bm{p_1}}{2}\, , \quad p_{N^*}
\approx p_5 + p_6 + p_3 \, ,
\end{equation}
For the wave function in momentum space $\tilde{\varphi}_{\bm{k}}( \bm{P} )$ 
we use an analytic expression obtained for a separable non-local potential 
that takes into account the repulsion~\cite{naqvi}. The parameters are 
chosen~\cite{naqvi} to fit the $^1S_0$ $n p$ phase shifts.       

\begin{figure}[h]
\begin{minipage}{.47\linewidth}
\centerline{\psfig{file=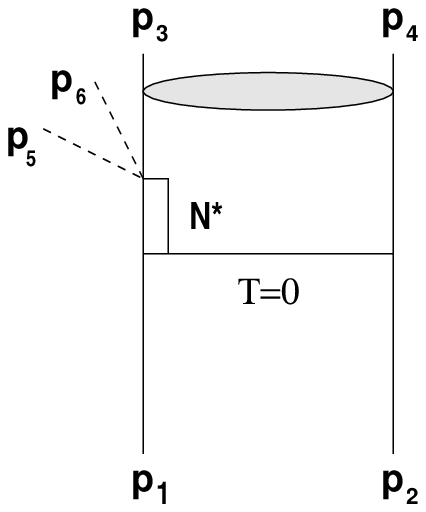,height=5.3cm,width=4.5cm}}
\end{minipage}
\hfill
\begin{minipage}{.47\linewidth}
\centerline{\psfig{file=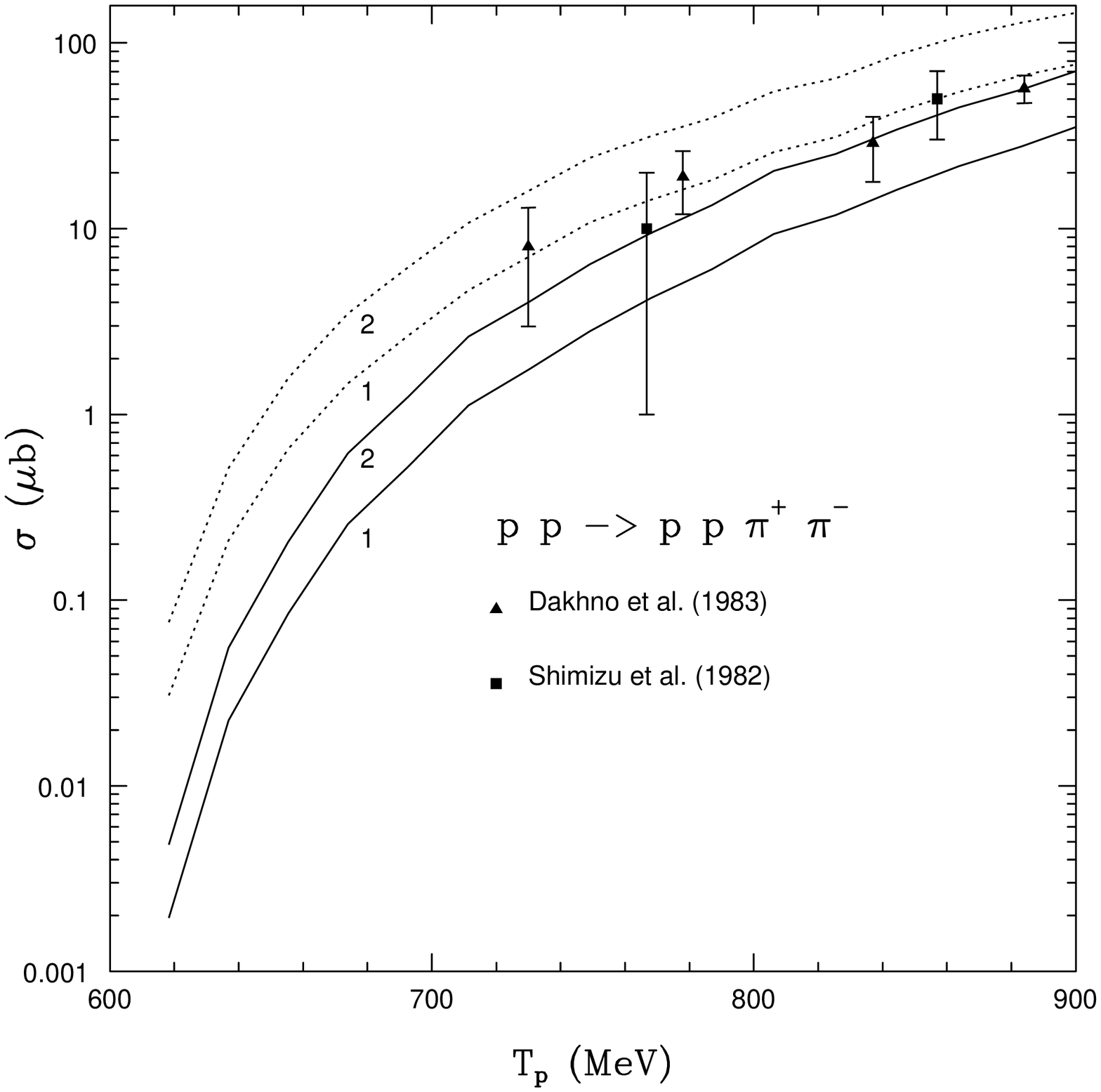,height=5.cm,width=5.cm}}
\end{minipage}
\caption{Dominant mechanism for $p p \rightarrow p p \pi^+ \pi^-$
at low energies with FSI included and total cross sections
obtained with ( dotted lines ) and without ( solid lines ) FSI for both 
sets I and II.}
\label{fig3}
\end{figure}

\section*{Acknowledgments}
This work has been partially supported by DGYCIT contract PB 96-0753.
L.A.R. acknowledges financial support from the Generalitat Valenciana.

\end{document}